\documentclass[
		,final
  ]
  {aipproc}

\layoutstyle{6x9}

\begin{document}

\title{Gravitino and Axino SuperWIMPs}


\author{J. A. R. Cembranos}{
  address={Department of Physics and Astronomy,
 University of California, Irvine, CA 92697 USA}
}

\author{Jonathan L.~Feng}{
  address={Department of Physics and Astronomy,
 University of California, Irvine, CA 92697 USA}
}

\author{A. Rajaraman}{
  address={Department of Physics and Astronomy,
 University of California, Irvine, CA 92697 USA}
}

\author{F. Takayama}{
  address={Institute for High-Energy Phenomenology, Cornell
University, Ithaca, NY 14853, USA} 
}

\begin{abstract}
Gravitinos and axinos produced in the late decays of other supersymmetric particles are well-motivated dark matter (DM) candidates, whose experimental evidences are very distinctive and different from other standard candidates, as thermal produced neutralinos in similar supersymmetric models. In particular, charged sleptons could appear stable because of the length of its lifetime. The direct production of such particles at both the Large Hadron Collider (LHC) and a future International Linear Collider (ILC) can give not only a clear signature of supersymmetry but also the first non-gravitational evidence of dark matter. 
\end{abstract}

\maketitle


\section{INTRODUCTION}

SuperWeakly-Interacting Massive Particles (superWIMPs) are as well motivated
dark matter candidates as the standard Weakly-Interacting Massive
Particles (WIMPs). Among the
most promising frameworks that incorporate superWIMPs are $R$-parity
conserving supersymmetry models in which the lightest supersymmetric
particle (LSP) is the gravitino or the axino. Future measurements of this scenario 
at the LHC or the ILC 
can predict important cosmological consequences to be tested 
with future improvements in astrophysical observations.

\section{SUPERWIMP PHENOMENOLOGY}

SuperWIMPs signatures have been analyzed from different points of
view~\cite{Feng:2003xh, SWIMPs}. On the one hand, this scenario modifies
light element abundances not only due to energy injections from 
late time decays~\cite{BBN} but also due to formation of new bound states~\cite{boundstates}. It has been argued that they can explain present inconsistent measurements, and alternatively, these analyses can be used to constraint superWIMP models, where the neutralino or a long lived slepton next LSP (NLSP) is seriously disfavored. On the other hand, late decays can also distort the Planck spectrum of the cosmic microwave background, which means that new experiments like DIMES can find evidence of these particles. Furthermore, sleptons NLSP produced in cosmic rays could be detected with high energy neutrino telescopes or in see water experiments \cite{Albuquerque:2003mi,Bi:2004ys}. 

\begin{figure}[bt]
  \includegraphics[height=7 cm
  ]{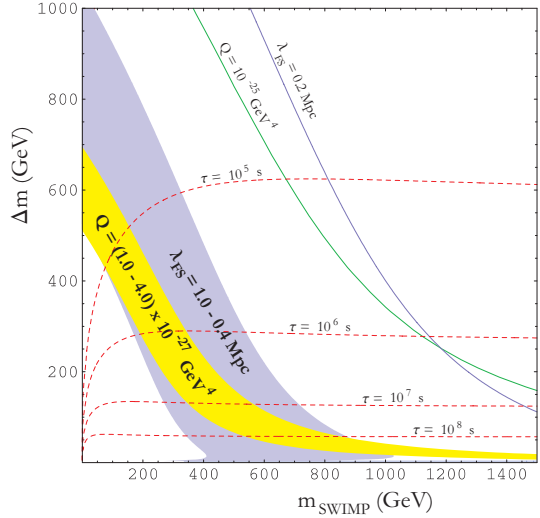}
\caption{Regions (shaded) of the $(m_{SWIMP}, \Delta m)$ plane
 preferred by small scale structure observations, where $\Delta m
 \equiv m_{NLSP} - m_{SWIMP}$, for gravitino superWIMP
 with photino NLSP.  The regions under both bands are disfavored.
 In the regions above both bands, superWIMP DM becomes similar to cold
 DM.  Contours of typical lifetimes $\tau$ are also
 shown~\cite{Cembranos:2005us}. }
\label{sph}
\end{figure}

Besides, the superWIMP behavior in relation to the structure formation of the Universe is, in general, different from the WIMP one.  Because superWIMP DM is
produced with large velocity in late decays, they can behave as Warm DM and resolve some observations which are not well understood in the Cold DM model. Indeed, a typical WIMP produces overdense cores in galactic halos, one or two order of magnitude more dwarf galaxies in the local Group than the observed one and disk galaxies with few angular momentum. 
All these disagreements can be fixed naturally in superWIMP scenarios as one can observe using the phase space density analysis and studying the damping in the power spectrum~\cite{Cembranos:2005us,Kaplinghat:2005sy,Sigurdson:2003vy} (see Figure \ref{sph}).

The collider signals of superWIMPs have attracted a lot of interest in the last years ~\cite{CollSW,Hamaguchi:2004df,Feng:2004yi}. 
The study of new particles will be a major component of the experimental programs at both the LHC and the ILC~\cite{Coll}. 
One of the most distinctive signals of superWIMP scenarios in colliders takes place when the NLSP is charged. Since its decay
can last for months or even years, it has been proposed that these charged particles 
can be trapped
outside of the particle physics detector so that their decays can be
studied.  A liquid trap that can be used to filter out the charged
particles so they can be transported to a quiet environment has been
purposed by Feng and Smith ~\cite{Feng:2004yi} while an active
stationary detector/trap has been prosed by Hamaguchi
et. al.~\cite{Hamaguchi:2004df}.  The liquid trap will be cost less
and allow for transportation to a clean, low background environment but
will be completely incapable of measuring the shorter range of the
lifetimes.  The active detector can measure almost the entire range of the
lifetimes, but will cost more and will have compensate for the
background from the high energy experiment and cosmic neutrinos.  Both
methods of trapping can be optimized with shaping and placing
the trap in the appropriate way to catch the maximum number of the
pseudo stable charged particles.  This is important because the LSP is
going to be identified by its decay distribution, not direct
detection.  It is imprint to catch as many of the charged particles
are possible so that an accurate identification of the superWIMP can be
made.  

\section{mSUGRA with SUPERWIMPs vs. mSUGRA with WIMPs}

SuperWIMP implications are completely different from those of WIMP.
 To illustrate it, we can consider minimal supergravity (mSUGRA)
-see, for example, \cite{Feng:2005ba}-. In the WIMP scenario, the slepton lightest
supersymmetric particle (LSP) region is excluded cosmologically.  In
the remaining region the neutralino is the LSP.  Much of the
neutralino LSP region is excluded because neutralinos are
overproduced, but some of this region is allowed.  In contrast, in the
superWIMP case, where, for instance, the gravitino or the axino is the
lightest supersymmetric particle (LSP), regions in which a slepton
is the lightest standard model superpartner are quit interesting, since
late decays to gravitinos or axinos can impact Big Bang
nucleosynthesis (BBN) and possibly even resolve some anomalies associated
with the production of $^6$Li and $^7$Li~\cite{Feng:2003xh, BBN, boundstates}. 
At the same time, much of the
neutralino region is disfavored since neutralinos typically have
two-body decays that produce hadrons, which destroy BBN successes. In
fact, in the neutralino case, the region excluded by overproduction
in the WIMP scenario are the most interesting in the superWIMP one,
since the abundance of the dark matter is reduced by the ratio of WIMP
to superWIMP masses when the WIMPs decay.

\section{CONCLUSIONS}

The fact that dark matter cannot be made of any of the known particles is one 
of the most arguments for the existence of new physics, be it in form of new particles
or as a modification of gravity at small energies \cite{Cembranos:2005fi}. The standard candidates for DM are WIMPs.  These emerge naturally in several
well-motivated particle physics frameworks, such as
supersymmetry~\cite{SUSY}, universal extra dimensions~\cite{UED} and
brane-worlds~\cite{BW1,BW2}. With masses and interactions at the
electro-weak scale, these are naturally produced with the correct DM
abundance. Furthermore, they behave as cold DM, which means that they
explain successfully the large scale structure of the Universe.

SuperWIMPs appear
naturally in the same well-motivated scenarios and their DM abundance
is also naturally the observed one, since they are produced in the
decays of WIMPs and naturally have similar masses. The astrophysical consequences 
and collider signatures of superWIMPs are, however, very
different from those of WIMPs. In these well-defined
particle models, astrophysical observations on small scale structures, 
distortions of the cosmic microwave background or light elements abundances
have direct implications
for possible measurements at future colliders.

\section{ACKNOWLEDGMENTS}

The work
of JARC and JLF is supported in part by NSF CAREER grant
No.~PHY--0239817, NASA Grant No.~NNG05GG44G, and the Alfred P.~Sloan
Foundation. The work of JARC is also supported the Fulbright-MEC program, 
and the FPA 2005-02327 project (DGICYT, Spain). The work of
AR is supported in part by NSF Grant No.~PHY--0354993.  The work of
FT is supported by the NSF grant No.~PHY-0355005.

\end{document}